\documentclass[aps,twocolumn,preprintnumbers,floats]{revtex4}
\usepackage{amsfonts}
\usepackage{amsmath}
\usepackage{graphicx}
\usepackage{dcolumn}
\usepackage{bm}

\begin{document}

\preprint{}
\title{Smallest 90$^o$ domains in epitaxial ferroelectric films}
\author{A.H.G. Vlooswijk$^{1}$, G. Catalan$^{2}$, A. Janssens$^{3}$,
B. Barcones$^{3}$, S. Venkatesan$^{4}$, G. Rijnders$^{3}$, B. Kooi$^{4}$, J.T.M de Hosson$^{4}$, D.H.A. Blank$^{3}$}
\author{B. Noheda$^{1}$}
\email{ b.noheda@rug.nl, Corresponding author}
\affiliation{$^{1}$Zernike Institute for Advanced Materials,
University of Groningen, Groningen 9747AG, The Netherlands}
\affiliation{$^{2}$Department of Earth Sciences, University of
Cambridge, Cambridge CB2 3EQ, United Kingdom}
\affiliation{$^{3}$MESA+ Institute for Nanotechnology, Twente
University, Enschede 7500 AE, The Netherlands}
\affiliation{$^{4}$ Dept. Applied Physics, Zernike Institute for Advanced Materials, Groningen 9747AG, The Netherlands}
\maketitle

\textbf{Ferroelectrics display spontaneous and switchable electrical polarization. Until recently, ferroelectricity was believed to disappear at the nanoscale; now, nano-ferroelectrics are being considered in numerous applications\cite{Scot07}. This renewed interest was partly fuelled by the observation of ferroelectric domains in films of a few unit cells thickness\cite{Ahn04,Fong04}, promising further size reduction of ferroelectric devices\cite{Greg}. It turns out that at reduced scales and dimensionalities the material's properties depend crucially on the intricacies of domain formation\cite{Bell}, that is, the way the crystal splits into regions with polarization oriented along the different energetically equivalent directions, typically at 180$^o$ and 90$^o$ from each other. Here we present a step forward in the manipulation and control of ferroelectric domains by the growth of thin films with regular self-patterned arrays of 90$^o$ domains only 7 nm wide. This is the narrowest width for 90$^o$ domains in epitaxial ferroelectrics that preserves the film lateral coherence, independently of the substrate.}

Nowadays, novel materials can be grown as very thin layers under strain, on an appropriate substrate. In ferroelectrics, in particular, strain can induce artificial symmetries and, potentially, new physical properties, as well as determine the domain configuration and/or the critical temperature\cite{Korn04,Dawb05,Bung04,Choi04,Pert98}. Recent results indicate that there is essentially no lower bound to the thickness at which a film can be ferroelectric, even when the polarization is perpendicular to the plane of the film\cite{Ahn04,Fong04}. This is thanks to the appearance of antiparallel ($180^{o}$) stripe domains, which minimize the ferroelectricity-suppressing effect of depolarization fields\cite{Meye02,Junq03}. Kittel's law\cite{Kitt49} also dictates that the domain size decreases as the square root of the film thickness, so the density of domains increases for very thin films. Since domains determine everything from coercive fields to
switching kinetics, electromechanical properties and dielectric response, the structure and behaviour of ferroelectric domains in the
nanoscale is rapidly becoming a very active area of research. Proper understanding of domains in very thin films is essential for harnessing the functional properties of ferroelectrics at the nanoscale and for further decreasing the size of ferroelectric devices\cite{Scot07}.

In the simplest and most common case of tetragonal thin films grown on cubic substrates, the orientation of the polarization in the
ferroelectric film depends on the relationship between the lattice parameters of the film (${a,c}$) and the substrate
(${a_s}$). If ${a_s\leq a}$, the film will grow with $c$ and, thus, the polarization, out-of-plane (\textit{c}-phase).
Such films would (if no or bad electrodes are present) split into 180$^{o}$ domains in order to decrease the depolarizing
fields\cite{Meye02,Junq03}, as mentioned above. These domains will form periodic stripe patterns\cite{Fong04}, similar to those found on
ferromagnetic thin films\cite{Kitt49}. In this scenario, the electrostatic energy is the sole responsible for domain formation.
Conversely, if $a_{s}\simeq c$, the film will grow with \textit{c} in-plane (\textit{a}-phase) and in-plane $a_{1}/a_{2}$ domains can form. Here only elastic considerations dictate the domain scaling\cite{Roit76,Pomp}.

For intermediate situations (${a< a_s< c}$) the film will usually consist of domains of \textit{c}- and \textit{a}-phase in a ratio such as to fit the substrate (see Fig.1a) and both elastic and depolarization effects compete with each other, as a function of thickness, temperature or electrode material\cite{Roit76,Pomp,Stem95,Fost96,Roel,Naga01,Pert96}. These c/a/c patterns are more difficult to engineer since they involve crystallographic tilts\cite{Shil} and since the strain required to stabilize them can also be accommodated by misfit dislocations and other defects, which themselves interfere with the domain pattern. Finding a right choice of film and substrate, which imposes enough strain without causing misfit dislocations is quite important but it was, until recently, difficult. Here we show that this is now possible by combining ferroelectric PbTiO$_3$ (PTO) with the recently introduced DyScO$_3$ (DSO), as a substrate\cite{Choi04,Bieg05}.

PTO is an archetype of tetragonal perovskite ferroelectrics (${a= 3.90\AA}$, ${c= 4.14 \AA}$), and also an end member of PZT -a material widely used in ferroelectric memories and electromechanical actuators. DyScO$_3$ (DSO) is an
orthorhombic perovskite with a pseudocubic lattice parameter of ${a_s \approx b_s=3.95 \AA}$\cite{Bieg05}. There is, however, a near-perfect lattice match of the two at the growth temperature, where PTO is in its cubic phase. This means that thin films of PTO can be grown with essentially zero strain, thus minimizing the appearance of defects. Only as the films cool down do strains develop and increase. At room temperature, the strain between tetragonal PTO and pseudocubic DSO is such that the elastic energy lies near the critical boundary for the appearance of \textit{a/c} domains\cite{Pomp}, and thus competition effects between depolarization and elastic energy are also easily observed. Indeed, as a function of thickness ($d$) we have observed a transition from depolarization-dominated, 180$^o$-domains ($d$= 5nm) to strain-dominated, 90$^o$-domains ($d$= 15, 30, 45, 60nm). We report that in these latter films the width of the a-domains is the smallest possible in order to keep the films' horizontal coherency, in PTO and any other perovskite ferroelectric.

The films of PTO were grown by pulsed laser deposition with thicknesses of 5, 15, 30, 45 and 60nm on (110)-oriented DyScO$_3$.
Some of the films were grown with an electrode SrRuO$_3$ buffer layer between the film and the substrate. Films with ${d}$= 5nm were
previously shown to be fully coherent and displayed 180$^{o}$ domains with the polarization tilted due to the epitaxial strain\cite{Cata06}. Here we show that at larger thicknesses, ($d\geq$ 15nm), the elastic energy developed during cooling is sufficient to induce $a$-domains. The reduced amount of defects allows the self-organization of these domains in very regular c/a/c patterns.

Figure 1b shows a synchrotron XRD area scan around the ${(100)_c}$ reflection. The intensity modulation, showing a periodicity
along the [H00]$_c$, direction is clearly visible. Scans around other reflections, both in-plane and out-of plane, have shown that
the modulation is present throughout reciprocal space with a spacing of ${\Delta H\approx 0.014}$, corresponding to a real-space
wavelength of ${\Lambda= 0.3950}$nm${/ \Delta H\ =}$28 nm. Similar satellites, with periodicities of ${29\pm 2}$ nm were observed in
all 30 nm thick films, with and without electrode bottom layer. The Bragg peak corresponding to the relaxed (bulk) \textit{c} lattice parameter of PbTiO$_3$ can be observed in the in-plane scan (at H= 0.952), showing the presence of \textit{a}-domains. Interestingly, a gradient of lattice parameters is observed in both \textit{a} and \textit{c}-domains\cite{Vloo07}.

Figure 2a shows an AFM image of a 28 nm thick film, similar to that used in Figure 1. It can be seen that the films grow as atomically
flat islands, or mounds\cite{Sanc04}, in agreement with the 2D electron diffraction patterns observed in-situ during growth. The figure shows narrow dip stripes running through the films, the width of which is below the resolution limit of the AFM tip (${\sim 9nm}$). Films grown on a SrRuO$_3$ electrode buffer layer show similar morphology and a Fourier analysis of one such film (see Figure 2b) reveals three different wavelengths of about 525, 150 and 30 nm. The first two correspond to the substrate terraces and to the size of the mounds, respectively. The third periodicity of 30nm corresponds to that of the dip vertical lines and agrees with that observed in the XRD patterns. By combining these results with TEM images (Figure 3,4) we can show that these lines are a grid of ultra-thin \textit{a}-domains running between larger c-domains. This morphology is a miniaturized version of that observed in thicker PZT layers\cite{Roel,Naga01}.

In Figure 3, a TEM image of one of the films shows a c/a/c pattern. The total periodicity ($\Lambda$) extracted
from the TEM images is 27 nm, in good agreement with the diffraction data of the same film (Figure 1). As explained, during cooling after the film growth, the mismatch between the $a$ lattice parameter of PTO and that of the pseudo-cubic (110)-DSO increases from 0, at the growth temperature, to $ \sim {1.3\%}$, at room temperature. The fraction of \textit{a}-domains needed to compensate this mismatch, in the absence of defects, can be simply estimated by imposing that ${N_c a + N_a c= Na_s}$ (Figure 1b), where ${N_a}$ and ${N_c}$ are the number of unit cells in the \textit{a}-domain and \textit{c}-domain, respectively, and $N$ is the total number of unit cells in one wavelength.
This leads to a fraction of unit cells in the \textit{a}-domain of ${N_a/N=(a_s - a)/(c - a)}$${ \approx 0.24}$, in
good agreement with the experimentally observed ratio. The quantity $(a_s - a)/(c -a)$ was defined in the seminal works by Pompe and
co-workers\cite{Pomp,Fost96} as the coherency strain ($e_{r}$). The stability limit for existence of \textit{a}-domains at large thicknesses predicted by these works is ${e_{r} \approx}$0.215, implying that PTO on DSO is close to the critical line that separates the stability of c/c (180$^o$) domains from a/c (90$^o$) domains. Moreover, the creation of \textit{a/c} domains requires an amount of elastic energy, whose density is proportional to the film thickness and, accordingly, the critical coherence strain is thickness-dependent. Following these authors, for PTO on DSO the transition from 180$^o$ to 90$^o$ domains should take place at around 10 nm, which is consistent with our observations.

One of the most interesting features of the TEM images is the strong tendency for the \textit{a}-domains to have the same width of about 7 nm. Constant or discretized (but bigger) widths of the \textit{a}-domains can be found in the literature\cite{Roel,Naga01,Ganp02,Fost96},
suggesting that they may be a general (but unexplained) feature of \textit{a/c} domains. High resolution TEM helps here to clarify their origin (see Figure 4a). The tilt angle, $\alpha$, between the \textit{a} and the \textit{c} domains, as measured from these images is 3-4$^o$, as expected for bulk (relaxed) PbTiO$_3$ from ${ \alpha= 90^o-2tan^{-1}(c/a)}$, which is the condition
for the \textit{a} and \textit{c} domains to be able to share the strain-free (101) plane at the domain wall\cite{Shil} (Figure 4b).
The large degree of crystallographic ordering of the unit cells in the \textit{c}-domains is in contrast
with the disorder observed within the \textit{a}-domains, in agreement with the broad XRD rocking curve observed for the
\textit{a}-domains (transversal [0K0]-scan at H= 0.952) in Fig. 1. Figure 4b shows that, except at the proximity of the interface
(where distortions are introduced by clamping to the substrate), the width of an \textit{a}-domain is such that the coherence of the
neighboring \textit{c}-domains is preserved. There is, therefore, a minimum possible size of the \textit{a}-domain in order to maintain
lateral coherence (collinearity of the atomic planes in adjacent \textit{c}-domains) that is ${w^{min}_{a}= c/{sin\alpha}}$.

 For PTO, $w^{min}_{a}{\sim 6.6 nm}$, in excellent agreement with the values observed in the TEM images. This analysis implies that,
in general, in a tetragonal film the width of the \textit{a}-domains must a multiple of $w^{min}_{a}$ in order to maintain the
horizontal coherence. Although, to the best of our knowledge, this has not been discussed so far, we have indeed found examples in the literature where $a$-domains with widths multiple of 7 nm are observed in PTO (see supplement)\cite{Fost96}. This conclusion is also valid for other materials and is independent of the substrate, as long as dislocations and other defects that can induce domain nucleation are avoided. For example, Pb(Zr$_{0.2}$Ti$_{0.8}$)O$_3$ has a=3.97 \AA and c=4.13 \AA\cite{Jano}; for films of this composition grown on $SrTiO_{3}$ substrates, a typical value of ${w}=$ 20nm was reported\cite{Naga01}, which is  $\sim 2w^{min}_{a}$, calculated for this material.

Since ${N_a/N= e_{r}}$, the existence of a minimum size for the \textit{a}-domains implies also the existence of a minimum size for the
domain wavelength. In our case, ${\Lambda =Na_s=}$ 27.5 nm for the domain periodicity, again in very good agreement with the one observed  in all the 30nm thick films (${\Lambda =}$27-32 nm). It is worth emphasizing that continuum theories such as Kittel's law\cite{Kitt49,Roit76} predict that the domain period scales as the square root of the film thickness. Kittel's law has been observed to work for epitaxial films with $180^{o}$ domains\cite{Fong04} and for free-standing films with $90^{o}$ domains\cite{Schi06}.
In our case an increase in the domain periodicity with increasing thickness (consistent with Kittel's law) is found for ${d\geq 30nm}$. However, for films of thicknesses ${d\leq 30nm}$, a larger domain wavelength has been observed, in agreement with the fact that ${\sim 27}$nm is indeed a lower bound for the a/c periodicity of epitaxial PTO on DSO.

For PTO on DSO and many other systems \cite{Fost96,Naga01,Roel,Ganp02}, the 90$^{o}$ a/c domains appear in
the form of wide \textit{c}-domains separated by narrow \textit{a}-stripes. The condition of horizontal coherency forces the \textit{a}-domains to adopt discrete sizes, which are substrate-independent and depend only on the tetragonality of the ferroelectric film. In the case of PbTi$O_3$ on DySc$O_3$, we have achieved the narrowest (7 nm) possible \textit{a}-domains, as well as the shortest \textit{c/a/c} domain periodicity compatible with such \textit{a}-domain size and the coherency strain. This also implies that epitaxial films with \textit{a/c} domains cannot continuously adjust their domain period to match Kittel's law, in contrast to the 180$^o$ domain scenario\cite{Fong04}. When the thickness is of the order of a few tens of nanometers, discrete jumps in periodicity as a function of thickness may be seen instead. Preserving lateral coherency leads to the enhancement of the registry of the self-patterned domain structures and it is of high interest in applications such as ferroelectric RAM memories.

The authors thank Arjen Molag, Wolfgang Caliebe, Gijsbert Rispens and Florencio S\'{a}nchez for useful discussions and Henk Bruinenberg for his technical assistance. Financial support from a Marie Curie Fellowship (G.C.) and the Dutch organizations FOM and NWO is gratefully acknowledged.

\textit{Methods}

The PbTiO$_3$ films and SrRuO$_3$ buffer layers were grown by Pulsed Laser Deposition assisted by \textit{in-situ} RHEED (Reflective High Energy Electron Diffraction). A pulsed excimer laser with $ \lambda$= 248 nm was used with a frequency of 1 Hz and a laser fluence of about 2 J/cm$^2$. A target-substrate distance of 45-48 mm was chosen. An O$_2$ pressure of 0.13 mbar and a substrate temperature of 570$^o$C were used during the deposition. After growth, the samples were cooled down in an O$_2$ atmosphere of 1 bar. The films were grown on (110)-oriented DyScO$_3$ substrates with nominal lattice parameters ${a_o}$= 5.44 \AA, ${b_o}$= 5.71 \AA and  ${b_o}$= 7.89 \AA, provided by CrysTec GmbH. In this orientation, the in-plane is a pseudo-square latttice (with orthogonal and very similar lattice vectors) with nominal ${a_{pc}= 3.945 \AA}$ and  ${b_{pc}=3.943 \AA}$. From our XRD measurement (without analiser crystal) we observe ${a_{pc}= b_{pc}= 3.95 \AA}$. X-ray diffraction was performed in standard reflection geometry as well as in grazing incidence geometry, both using the CuK$\alpha$  radiation of a PANalytical X'pert diffractometer and using synchrotron radiation at the wiggler W1 beamline in Hasylab(DESY-Hamburg). The atomic force microscopy images have been taken with a Nanoscope III microscope using BudgetSensors Tap300Al tapping mode tips with an Aluminium reflex coating and a tip radius below 10nm. The AFM data were analysed using the WSxM free software downloadable at http://www.nanotec.es. The TEM images presented here were taken with a 300 kV Philips CM300ST - FEG Microscope.

\begin{widetext}

\begin{figure}
\includegraphics[scale=0.7]{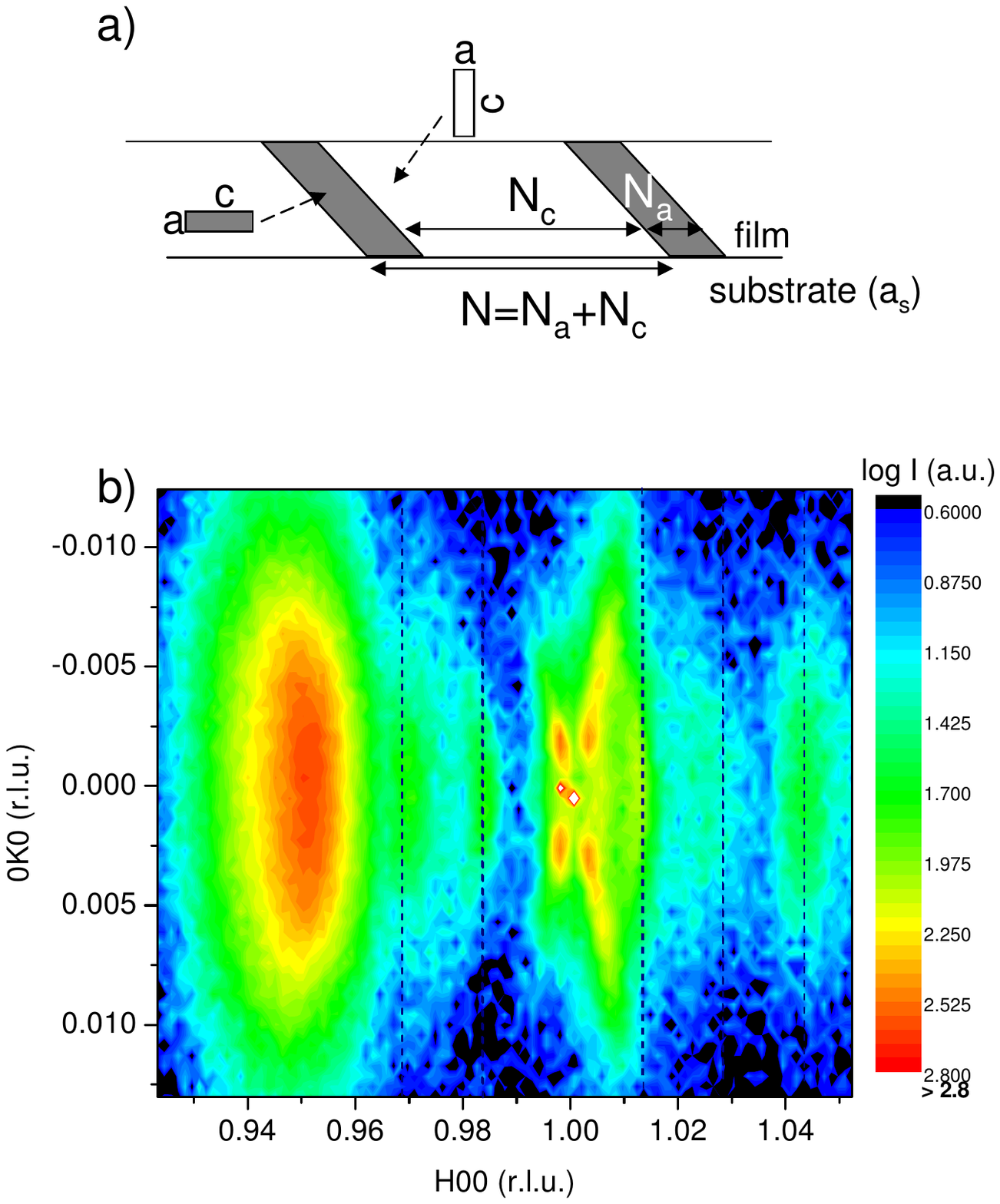}

\caption{(a) Sketch of a c/a/c domain pattern for a tetragonal film with ${a< a_s< c}$. a, c are the lattice parameters of the film and $a_s$ is that of the substrate. (b) c/a/c domain periodicity by synchrotron x-ray diffraction. Reciprocal space area scans in the in-plane (HK0) scattering plane, around the ${(100)_c}$ reflection, by synchrotron x-ray diffraction in grazing indicence geometry. In this particular sample, a SrRuO$_3$ electrode layer was deposited and the orthorhombic four-fold twinning of this buffer layer can be observed\protect\cite{Twen}. The axes units are reciprocal lattice units of the substrate in pseudo-cubic notation (${1 r.l.u= 2\pi/3.95 \AA}$). The wavelength is ${\lambda= 1.26515\AA}$. The intensities are plotted in logarithmic scale (from low to high: blue-green-yellow-red-white)}
\end{figure}

\begin{figure}
\includegraphics[scale=0.7]{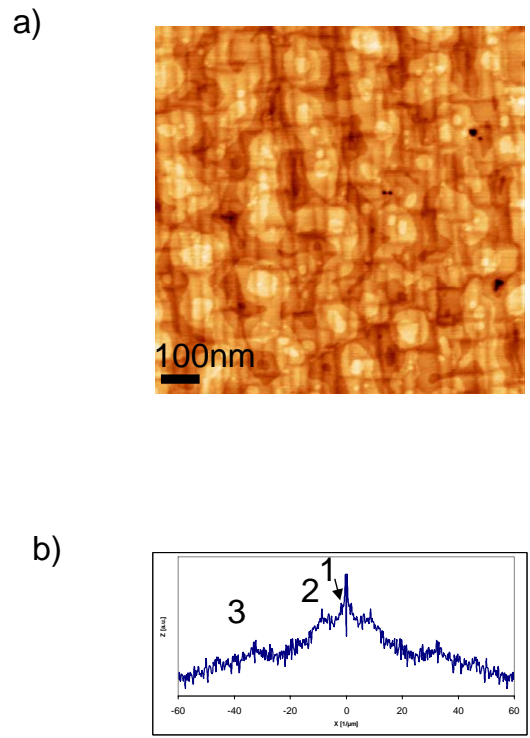}
\caption{(a) c/a/c/ domain periodicity by atomic force microscopy. AFM topography image of a 28nm thick PbTiO$_3$ film on DyScO$_3$ showing a RMS roughness of 3.7 ${\AA}$ (full z-scale is 3nm).
(b) Horizontal line scan of the Fast Fourier Transform of a larger 4x4${\mu}$m AFM image of a similar 28nm thick PbTiO$_3$ film on DyScO$_3$ with a SrRuO$_3$ buffer layer. No difference is found in films with and without the electrode layer. Three periodic features can be distinguished and linked to the topography image: 1) is a periodicity of 525nm, along a single direction only. This is the faint footprint of the terrace steps of the substrate; 2) represents the size uniformity of the atomically-flat growth islands; 3) reveals a 31 nm periodicity. In combination with the TEM data, this one can be assigned to the ferroelastic-ferroelectric c/a/c domain pattern. Image (a) clearly shows that this domain pattern is superimposed onto the island morphology, suggesting that the a/c domains are unaffected by the growth mode, and that the film is indeed continuous under the surface mesas.}
\end{figure}

\begin{figure}
\includegraphics[scale=0.7]{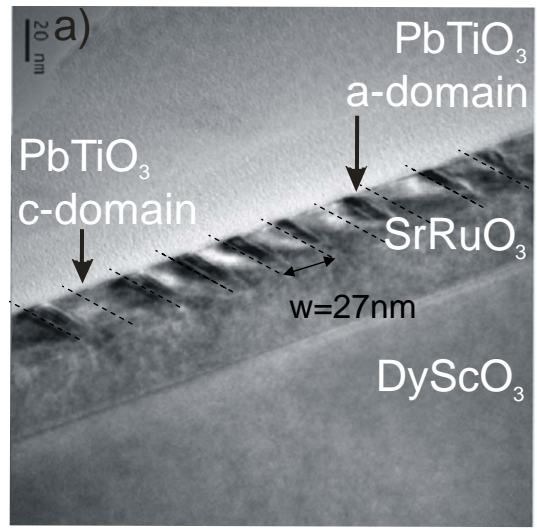}

\caption{c/a/c domain periodicity by Transmission Electron Microscopy. TEM image of a 30nm thick film of PbTiO$_3$ on a SrRuO$_3$ buffer layer on DyScO$_3$, the same film used for Figure 1. The tilt of the specimen was chosen to increase the contrast between the a- and c-domains. The image shows the c/a/c domain patterns and the uniform size of the a-domain bands. The size of the c-domains shows a somewhat larger variation with an average total period of 27 nm. Often a double c-domain is found, as also seen in this image. The dashed lines signal the perfect periodicity. The observation of modulations in the XRD patterns (see Figure 1b) shows that this periodicity, despite local disruptions, is robust throughout the film, as corroborated in different TEM and AFM images.}
\end{figure}

\begin{figure}[tbp]
\includegraphics[scale=0.7]{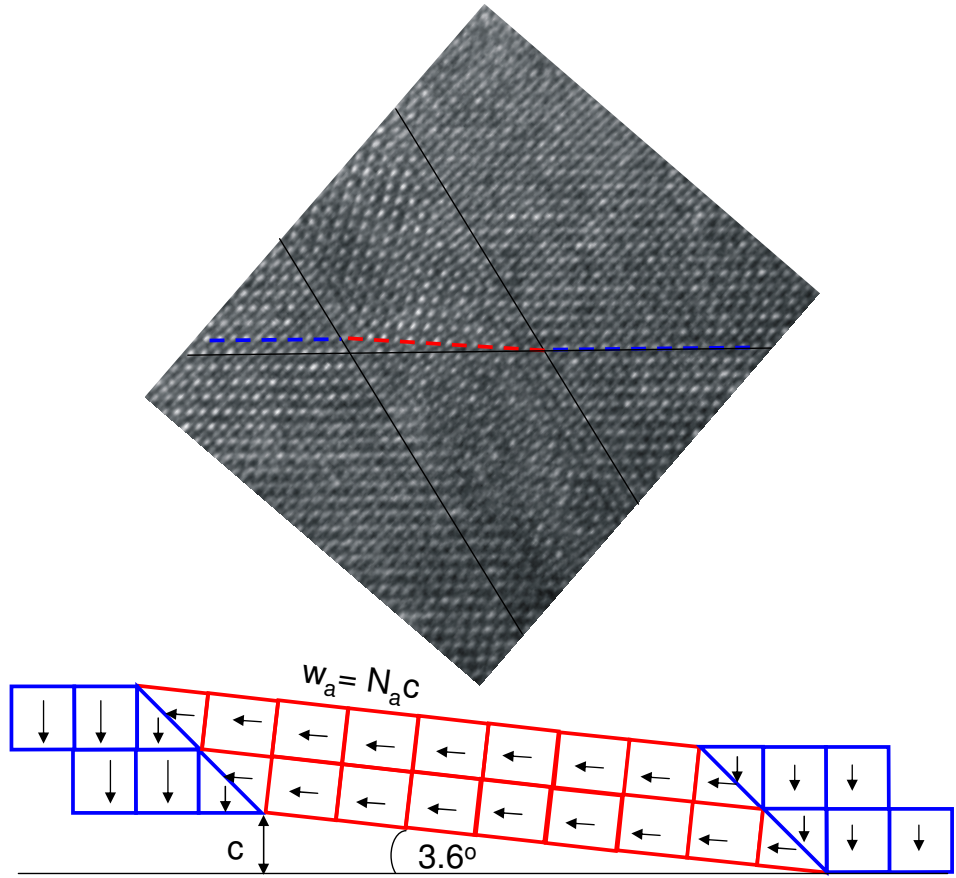}
	\caption{a) HRTEM image of a 30nm film of PbTiO$_3$ on a SrRuO$_3$ buffer layer on a DyScO$_3$ substrate,
around an $a$-domain band. The dashed lines show how the coherent connection between two c-domains is mediated by the narrow $a$-domain;
b) sketch showing the coherent horizontal match of the c/a/c chain, the tetragonal
tilts at the two boundaries, and how this determines the minimum possible width of the
a-domain (${w_a}$). Note that close to the interface, where the film has to attach to the (flat) substrate, this picture does not hold. There, the strains at the interface due to the clamping of the a-domain will tend to reduce the width of this domain in a way that will depend on the elasticity of the substrate. This also introduces a large degree of disorder in the $a$-domains compared to that of the c-domains.}
\end{figure}

\begin{figure}[tbp]
\includegraphics[scale=0.7]{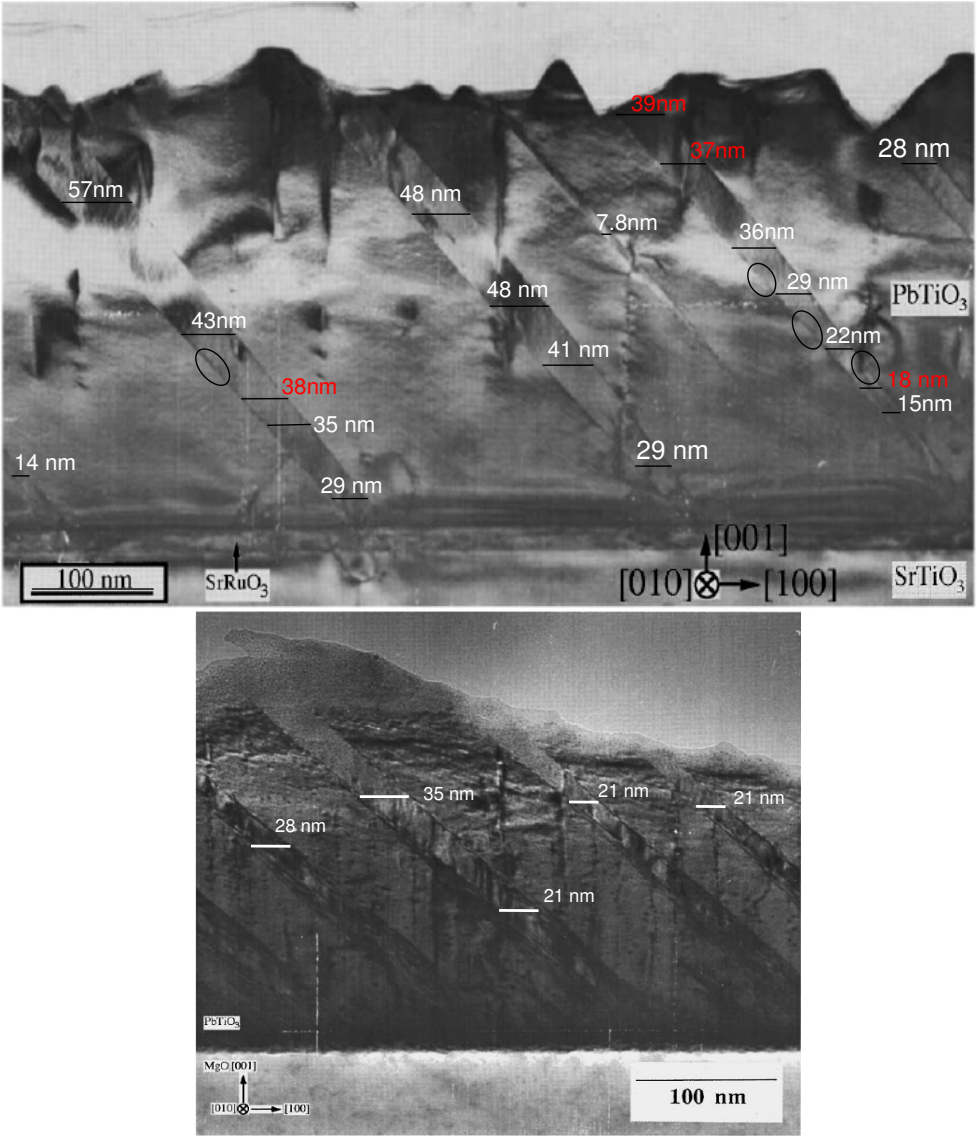}
\caption{\textit{Supplementary Figure 1} One can find in the literature other examples confirming our observations.
In this figure we include TEM images taken from Foster et al, Journal of Appl. Phys. 79, 1405 (1996), of a thick
PbTiO$_3$ film deposited on a SrTiO$_3$ substrate. Most of the a-domains are found to have widths that
are multiples of 7 nm (${\pm1}$) nm  (the ones that are found not to fulfill this are indicated in red).
In the top image, sligtly wedged-shaped domains can be observed. In these cases it is sometimes
possible to see discrete jumps in domain width (encircled in the image). Both the a-domains and the domain
periodicity are larger than in our films due to the larger thickness of these films.}
\end{figure}

\end{widetext}
\end{document}